\documentclass[traditabstract,final]{aa}
\usepackage{amsmath}
\usepackage{txfonts}
\usepackage{alltt}
\usepackage{ifpdf}

\ifpdf
\usepackage[pdftex]{graphicx}
\usepackage{epstopdf}
\usepackage[pdftex,colorlinks=false,pdfstartview=Fit,urlcolor=pdfwww,
  bookmarks=true,bookmarksnumbered=true]{hyperref}

\else
\usepackage[dvips]{graphicx}
\usepackage[dvips]{hyperref}

\fi
\usepackage{url}
\usepackage{framed}
\usepackage{natbib,ifthen}

\bibpunct{(}{)}{;}{a}{}{,} 

\newcommand{\deleted}[1]{{\bf\ (DELETED TEXT)}}

\newcommand{\lmax}{\ensuremath{l_{\text{max}}}}
\newcommand{\nside}{\ensuremath{N_{\text{side}}}}

\newcommand{\alm}{\ensuremath{a_{lm}}}
\newcommand{\libpsht}{\texttt{libpsht}}
\newcommand{\Libpsht}{\texttt{Libpsht}}
\newcommand{\libsharp}{\texttt{libsharp}}
\newcommand{\Libsharp}{\texttt{Libsharp}}
\newcommand{\shtns}{\texttt{shtns}}

\newcommand{\stwohat}{S${}^2$HAT}
\newcommand{\nifty}{NIFT\textsc{y}}
\newcommand{\lamlm}{\ensuremath{{{}_s\lambda_{lm}(\vartheta)}}}
\newcommand{\fnurl}[1]{\footnote{\url{#1}}}
\newcommand{\compl}[1]{\ensuremath{\mathcal{O}(#1)}}

\newcommand{\pnull}{\phantom{0}}

\begin{document}

\title {Libsharp -- spherical harmonic transforms revisited}
\author {M.~Reinecke\inst{1} \and
 D.~S.~Seljebotn\inst{2}}
\institute {Max-Planck-Institut f\"ur Astrophysik, Karl-Schwarzschild-Str.~1, 85741 Garching, Germany \\ \email{martin@mpa-garching.mpg.de}\and
Institute of Theoretical Astrophysics, University of Oslo, P.O.\ Box 1029 Blindern, N-0315 Oslo, Norway \\ \email{d.s.seljebotn@astro.uio.no}}


\date{Received 18 March 2013 / Accepted 14 April 2013}

\abstract {
We present \libsharp, a code library for spherical harmonic transforms (SHTs),
which evolved from the \libpsht\ library and addresses
several of its shortcomings, such as adding MPI support for distributed memory
systems and SHTs of fields with arbitrary spin, but also supporting new developments in
CPU instruction sets like the Advanced Vector Extensions (AVX) or fused
multiply-accumulate (FMA) instructions. The library is
implemented in portable C99 and provides an interface that can be easily accessed from
other programming languages such as C++, Fortran, Python, etc. Generally, \libsharp's
performance is at least on par with that of its predecessor; however, significant
improvements were made to the algorithms for scalar SHTs, which are roughly
twice as fast when using the same CPU capabilities. The library is available
at \url{http://sourceforge.net/projects/libsharp/} under the terms of the
GNU General Public License.
\keywords {methods: numerical -- cosmic background radiation -- large-scale structure of the Universe} }

\maketitle

\section{Motivation}
While the original \libpsht\ library presented by \citet{reinecke-2011}
fulfilled most requirements on an implementation of spherical harmonic
transforms (SHTs) in the astrophysical
context at the time, it still left several points unaddressed. Some of those
were already mentioned in the original publication: support for SHTs
of arbitrary spins and parallelisation on computers with distributed memory.

Both of these features have been added to \libpsht\ in the meantime, but other,
more technical, shortcomings of the library have become obvious since its
publication, which could not be fixed within the \libpsht\ framework.

One of these complications is that the library design did not anticipate
the rapid evolution of microprocessors during the past few years. While the code
supports both traditional scalar arithmetic as well as SSE2 instructions,
adding support for the newly released Advanced Vector Extensions (AVX) and
fused multiply-accumulate instructions (FMA3/FMA4) would require adding
significant amounts of new code to the library, which is inconvenient and very
likely to become a maintenance burden in the long run. Using proper abstraction
techniques, adding a new set of CPU instructions could be achieved by only very
small changes to the code, but the need for this was unfortunately not anticipated
when \libpsht\ was written.

Also, several new, highly efficient SHT implementations have been published
in the meantime; most notably Wavemoth \citep{seljebotn-2012} and
\shtns\ \citep{schaeffer-2013}. These codes demonstrate that \libpsht's computational
core did not make the best possible use of the available CPU resources.
Note that Wavemoth is currently an experimental research code not meant
for general use.

To address both of these concerns, the library was redesigned from scratch.
The internal changes also led to a small loss of functionality; the new code
no longer supports multiple simultaneous SHTs of different type
(i.e.\ having different directions or different spins).
Simultaneous transforms of identical type are still available, however.

As a fortunate consequence of this slight reduction in functionality,
the user interface could be simplified dramatically, which is especially helpful
when interfacing the library with other programming languages.

Since backward compatibility is lost, the new name \libsharp\ was
chosen for the resulting code, as a shorthand for
``Spherical HARmonic Package''.

The decision to develop \libsharp\ instead of simply using
\shtns\ was taken for various reasons: \shtns\ does not support
spin SHTs or allow MPI parallelisation, it requires more main memory
than \libsharp, which can be problematic for high-resolution runs, and
it relies on the presence of the FFTW library. Also, \shtns\
uses a syntax for expressing SIMD operations,
which is currently only supported by the \texttt{gcc} and \texttt{clang} compilers,
thereby limiting its portability at least for the vectorised
version. Finally, \libsharp\ has support for partial spherical
coverage and a wide variety of spherical grids, including
Gauss-Legendre, ECP, and HEALPix.

\section{Problem definition}
This section contains a quick recapitulation of equations presented in
\citet{reinecke-2011}, for easier reference.

We assume a spherical grid with $N_\vartheta$ iso-latitude rings (indexed
by $y$). Each of these in turn consists of $N_{\varphi,y}$ pixels (indexed
by $x$), which are equally spaced in $\varphi$, the azimuth of the first ring
pixel being $\varphi_{0,y}$.

A continuous spin-$s$ function defined on the sphere with a spectral band limit
of \lmax\ can be represented either as a set of spherical harmonic coefficients
$_sa_{lm}$, or a set of pixels ${\vec p}_{xy}$.
These two representations are connected by spherical harmonic synthesis
(or backward SHT):
\begin{equation}
{\vec p}_{xy}= \sum_{m=-\lmax}^{\lmax} \sum_{l=|m|}^{\lmax} {}_sa_{lm}\, {}_s\lambda_{lm}({\vec\vartheta}_y)
 \exp{\left(im\varphi_{0,y} + \frac{2\pi imx}{N_{\varphi,y}}\right)}
\label{syn_used}
\end{equation}
and spherical harmonic analysis (or forward SHT):
\begin{equation}
{}_s{\hat a}_{lm} = \sum_{y=0}^{N_\vartheta-1} \sum_{x=0}^{N_{\varphi,y}-1} {\vec p}_{xy}\, {\vec w}_y\, {}_s\lambda_{lm}({\vec\vartheta}_y)\exp{\left(-im\varphi_{0,y} - \frac{2\pi i m x}{N_{\varphi,y}}\right)} \text{,}
\label{ana_used}
\end{equation}
where $\lamlm:={}_sY_{lm}(\vartheta,0)$ and ${\vec w}_y$ are quadrature weights.

Both transforms can be subdivided into two stages:
\begin{align}
{\vec p}_{xy}&=\sum_{m=-\lmax}^{\lmax}F_{m,y}\exp{\left(im\varphi_{0,y} + \frac{2\pi imx}{N_{\varphi,y}}\right)}
\quad\text{with}
\label{syn1}
\\
F_{m,y}&:=\sum_{l=|m|}^{\lmax} {}_sa_{lm}\, {}_s\lambda_{lm}({\vec\vartheta}_y)\text{,\quad and}
\label{syn2}
\\
{}_s{\hat a}_{lm} &= \sum_{y=0}^{N_\vartheta-1}G_{m,y}\,{}_s\lambda_{lm}({\vec\vartheta}_y)\quad\text{with}
\label{ana1}
\\
G_{m,y} &:= {\vec w}_y\sum_{x=0}^{N_{\varphi,y}-1} {\vec p}_{xy}\, \exp{\left(-im\varphi_{0,y}-\frac{2\pi i m x}{N_{\varphi,y}}\right)}\text{.}
\label{ana2}
\end{align}
Eqs.\ \ref{syn1} and \ref{ana2} can be computed using fast Fourier transforms
(FFTs), while Eqs.\ \ref{syn2} and \ref{ana1}, which represent the bulk of the
computational load, are the main target for optimised implementation in
\libsharp.

\section{Technical improvements}
\subsection{General remarks}
\label{remarks}

As the implementation language for the new library, ISO C99 was chosen. This
version of the C language standard is more flexible than the C89 one adopted
for \libpsht\ and has gained ubiquitous compiler support by now.
Most notably, C99 allows definition of new variables anywhere in the
code, which improves readability and eliminates a common source of programming
mistakes. It also provides native data types for complex numbers, which allows
for a more concise notation in many places. However, special care must be taken
not to make use of these data types in the library's public interface, since
this would prevent interoperability with C++ codes (because C++ has a different
approach to complex number support). Fortunately, this drawback can be worked
around fairly easily.

A new approach was required for dealing with the growing variety of instruction
sets for arithmetic operations, such as traditional scalar instructions, SSE2
and AVX. Rewriting the library core for each of these alternatives would be
cumbersome and error-prone. Instead we introduced the concept of a generic
``vector'' type containing a number of double-precision IEEE values and defined
a set of abstract operations (basic arithmetics, negation, absolute value,
multiply-accumulate, min/max, comparison, masking and blending) for this type. Depending
on the concrete instruction set used when compiling the code, these operations
are then expressed by means of the appropriate operators and intrinsic function
calls. The only constraint on the number of values in the vector type is that
it has to be a multiple of the underlying instruction set's native vector length
(1 for scalar arithmetic, 2 for SSE2, 4 for AVX).

Using this technique, adding support for new vector instruction sets is
straightforward and carries little risk of breaking existing code.
As a concrete example, support for the FMA4 instructions present in AMD's
Bulldozer CPUs was added and successfully tested in less than an hour.

\subsection{Improved loop structure}

\begin{figure}
\begin{framed}
\begin{alltt}
for b = <all submaps or "blocks">
  for m = [0;mmax]                       // OpenMP
    for l = [m;lmax]
      precompute recursion coefficients
    end l
    for y = <all rings in submap b>      // SSE/AVX
      for l = [m;lmax]
        compute s_lambda_lm(theta_y)
        for j = <all jobs>
          accumulate F(m,theta_y,j)
        end j
      end l
    end y
  end m

  for y = <all rings in submap b>        // OpenMP
    for j = <all jobs>
      compute map(x,y,j) using FFT
    end j
  end y
end b
\end{alltt}\end{framed}
\caption{Schematic loop structure of \libsharp's shared-memory synthesis code.}
\label{a2msimple}
\end{figure}

\begin{figure}
\begin{framed}
\begin{alltt}
for b = <all submaps or "blocks">
  for y = <all rings in submap b>        // OpenMP
    for j = <all jobs>
      compute G(m,theta_y,j) using FFT
    end j
  end y

  for m = [0;mmax]                       // OpenMP
    for l = [m;lmax]
      precompute recursion coefficients
    end l
    for y = <all rings in submap b>      // SSE/AVX
      for l = [m;lmax]
        compute s_lambda_lm(theta_y)
        for j = <all jobs>
          compute contribution to a_lm(j)
        end j
      end l
    end y
  end m
end b
\end{alltt}
\end{framed}
\caption{Schematic loop structure of \libsharp's shared-memory analysis code.}
\label{m2asimple}
\end{figure}

After publication of SHT implementations, which perform significantly better
than \libpsht, especially for $s=0$ transforms
\citep{seljebotn-2012,schaeffer-2013}, it became obvious that some bottleneck
must be present in \libpsht's implementation. This was identified with
\libpsht's approach of first computing a whole $l$-vector of \lamlm\ in
one go, storing it to main memory, and afterwards re-reading it sequentially
whenever needed. While the $l$-vectors are small enough to fit into the CPU's
Level-1 cache, the store and load operations nevertheless caused some
latency. For $s=0$ transforms with their comparatively low arithmetic operation
count (compared to the amount of memory accesses), this
latency could not be hidden behind floating point operations and so resulted
in a slow-down. This is the most likely explanation for the observation that
\libpsht's $s=0$ SHTs have a much lower FLOP rate compared to those with
$s\neq0$.

It is possible to avoid the store/load overhead for the \lamlm\ by applying
each value immediately after it has been computed, and discarding it as soon
as it is not needed any more. This approach is reflected in the loop structure
shown in Figs.~\ref{a2msimple} and \ref{m2asimple}, which differs from the
one in \citet{reinecke-2011} mainly by the fusion of the central loops over
$l$.

In this context another question must be addressed: the loops marked as
``SSE/AVX'' in both figures are meant to be executed in ``blocks'', i.e.
by processing several $y$ indices simultaneously. The block size is equivalent to
the size of the generic vector type described in Sect.~\ref{remarks}.
The best value for this parameter depends on hardware characteristics of the
underlying computer and therefore cannot be
determined a priori. \Libsharp\ always uses a multiple of the
system's natural vector length and estimates the best value by running quick
measurements whenever a specific SHT is invoked for the first time.
This auto-tuning step approximately takes a tenth of a wall-clock second.

Due to the changed central loop of the SHT implementation, it is no longer
straightforward to support multiple simultaneous transforms with differing
spins and/or directions, as \libpsht\ did -- this would lead to a combinatorial
explosion of loop bodies that have to be implemented. Consequently, \libsharp,
while still supporting simultaneous SHTs, restricts them to have the same spin
and direction. Fortunately, this is a very common case in many application
scenarios.

\subsection{Polar optimisation}
\label{polaropt}
As previously mentioned in \citet{reinecke-2011}, much CPU time can be saved by
simply not computing terms in Eqs.\ \eqref{syn2} and \eqref{ana1} for which
${}_s\lambda_{lm}(\vartheta)$ is so small that their contribution to the result
lies well below the numerical accuracy. Since this situation occurs for rings
lying close to the poles and high values of $m$, \citet{schaeffer-2013} referred
to it as ``polar optimisation''.

To determine which terms can be omitted, \libsharp\ uses the approach described
in \citet{prezeau-reinecke-2010}. In short, all terms for which
\begin{equation}
\label{eq_polaropt}
\sqrt{m^2+s^2-2ms\cos\vartheta}-\lmax \sin\vartheta > T
\end{equation}
are skipped. The parameter $T$ is tunable and determines the overall accuracy
of the result. \Libsharp\ models it as
\begin{equation}
T=\max(100,0.01\lmax)\text{,}
\end{equation}
which has been verified to produce results equivalent to those of SHTs without
polar optimisation.

\section {Added functionality}

\subsection{SHTs with arbitrary spin}
\label{spinsht}

While the most widely used SHTs in cosmology are performed on quantities
of spins 0 and 2 (i.e.\ sky maps of Stokes I and Q/U parameters), there is also
a need for transforms at other spins. Lensing computations require SHTs of
spin-1 and spin-3 quantities (see, e.g., \citealt{lewis-2005}).
The most important motivation for high-spin SHTs, however, are all-sky convolution
codes (e.g.\ \citealt{wandelt-gorski-2001,prezeau-reinecke-2010}) and
deconvolution map-makers (e.g.\ \citealt{keihanen-reinecke-2012}). The
computational cost of these algorithms is dominated by calculating expressions
of the form
\begin{equation}
R_{mk}(\vartheta) = \sum_{l=\max(|m|,|k|)}^{\lmax}
 a_{lm}b^{\ast}_{lk}d^l_{mk}(\vartheta)\text{,}
\end{equation}
where $a$ and $b$ denote two sets of spherical harmonic coefficients (typically
of the sky and a beam pattern) and $d$ are the Wigner $d$ matrix elements.
These expressions can be interpreted and solved efficiently
as a set of (slightly modified) SHTs with spins ranging from 0 to
$k_\text{max}\le \lmax$, which in today's applications can take on values much
higher than 2.

As was discussed in \citet{reinecke-2011}, the algorithms used by
\libpsht\ for spin-1 and spin-2 SHTs become inefficient and inaccurate for
higher spins. To support such transforms in \libsharp,
another approach was therefore implemented, which is based on the recursion
for Wigner $d$ matrix elements presented in \citet{prezeau-reinecke-2010}.

Generally, the spin-weighted spherical harmonics are related to the Wigner $d$
matrix elements via
\begin{equation}
{}_s\lambda_{lm}(\vartheta)= (-1)^m \sqrt{\frac{2l+1}{4\pi}}d^l_{-ms}(\vartheta)
\end{equation}
\citep{goldberg-etal-1967}. It is possible to compute the $d^l_{mm'}(\vartheta)$ using a three-term
recursion in $l$ very similar to that for the scalar $Y_{lm}(\vartheta)$:
\begin{align}
&\left[ \cos\vartheta - \frac{mm^\prime}{l(l+1)}\right] d^{l}_{mm^\prime}(\vartheta) \nonumber \\
=& \frac{\sqrt{(l^2-m^2)(l^2-{m^\prime}^2)}}{l(2l+1)}d^{l-1}_{mm^\prime}(\vartheta) \nonumber \\
+&\frac{\sqrt{[(l+1)^2-m^2][(l+1)^2-{m^\prime}^2]}}{(l+1)(2l+1)}d^{l+1}_{mm^\prime}(\vartheta)
\end{align}
\citep{kostelec-rockmore-2008}. The terms depending only on $l$, $m$, and $m'$
can be re-used for different colatitudes, so that the real cost of a recursion
step is two additions and three multiplications.

In contrast to the statements made in \citet{mcewen-wiaux-2011}, this recursion
is numerically stable when performed in the direction of increasing
$l$; see, e.g., Sect.\ \ref{acctest} for a practical confirmation.
It is necessary, however, to use a digital floating-point representation with
enhanced exponent range to avoid underflow during the recursion, as is
discussed in some detail in \citet{prezeau-reinecke-2010}.

\subsection{Distributed memory parallelisation}

\begin{figure}
\begin{framed}
\begin{alltt}
for m = <all local m>                // OpenMP
  for l = [m;lmax]
    precompute recursion coefficients
  end l
  for y = <all rings in the map>     // SSE/AVX
    for l = [m;lmax]
      compute s_lambda_lm(theta_y)
      for j = <all jobs>
        accumulate F(m,theta_y,j)
      end j
    end l
  end y
end m

rearrange F(m,theta_y,j) among tasks // MPI

for y = <all local rings>            // OpenMP
  for j = <all jobs>
    compute map(x,y,j) using FFT
  end j
end y
\end{alltt}
\end{framed}
\caption{Schematic loop structure of \libsharp's distributed-memory synthesis code.}
\label{a2mdist}
\end{figure}

\begin{figure}
\begin{framed}
\begin{alltt}
for y = <all local rings>            // OpenMP
  for j = <all jobs>
    compute G(m,theta_y,j) using FFT
  end j
end y

rearrange G(m,theta_y,j) among tasks // MPI

for m = <all local m>                // OpenMP
  for l = [m;lmax]
    precompute recursion coefficients
  end l
  for y = <all rings in the map>     // SSE/AVX
    for l = [m;lmax]
      compute s_lambda_lm(theta_y)
      for j = <all jobs>
        compute contribution to a_lm(j)
      end j
    end l
  end y
end m
\end{alltt}
\end{framed}
\caption{Schematic loop structure of \libsharp's distributed-memory analysis code.}
\label{m2adist}
\end{figure}

When considering that, in current research, the required band limit for
SHTs practically never exceeds $\lmax=10^4$, it seems at first glance
unnecessary to provide an implementation supporting multiple nodes.
Such SHTs fit easily into the memory of a single typical compute node and
are carried out within a few seconds of wall clock time. The need for
additional parallelisation becomes apparent, however, as soon as the SHT is no
longer considered in isolation, but as a (potentially small) part of another
algorithm, which is \libsharp's main usage scenario. In such a situation, large
amounts of memory may be occupied by data sets unrelated to the SHT, therefore
requiring distribution over multiple nodes.
Moreover, there is sometimes the need for very many SHTs in sequence, e.g.\ if they
are part of a sampling process or an iterative solver. Here, the parallelisation to a very large
number of CPUs may be the only way of reducing the time-to-solution
to acceptable levels. Illustrative examples for this are the Commander code
\citep{eriksen-etal-2008} and the \texttt{artDeco} deconvolution mapmaker
\citep{keihanen-reinecke-2012}; for the processing of high-resolution
\textit{Planck} data, the latter is expected to require over 100GB of memory and
several hundred CPU cores.

\Libsharp\ provides an interface that allows collective execution of SHTs
on multiple machines with distributed memory. It makes use of the
MPI\fnurl{http://en.wikipedia.org/wiki/Message_Passing_Interface} interface
to perform the necessary inter-process communication.

In contrast to the standard, shared-memory algorithms, it is the responsibility
of the library user to distribute map data and \alm\ over the individual
computers in a way that ensures proper load balancing. A very straightforward
and reliable way to achieve this is a ``round robin'' strategy: assuming $N$
computing nodes, the map is distributed such that node $i$ hosts the map rings
$i$, $i+N$, $i+2N$, etc.\ (and their counterparts on the other hemisphere).
Similarly, for the spherical harmonic coefficients, node $i$ would hold all
\alm\ for $m=i$, $i+N$, $i+2N$, etc.
Other efficient distribution strategies do of course exist and may be
advantageous, depending on the circumstances under which \libsharp\ is called;
the library makes no restrictions in this respect.

The SHT algorithm for distributed memory architectures is analogous
to the one used in the \stwohat\ package\fnurl{http://code.google.com/p/s2hat-library/}
and first published in \citet{szydlarski-etal-2013}; its structure is sketched
in Figs.\ \ref{a2mdist} and \ref{m2adist}.
In addition to the \stwohat\ implementation, the SHT will be broken down into
smaller chunks if the average number of map rings per MPI task exceeds a certain
threshold. This is analogous to the use of chunks in the scalar and
OpenMP-parallel implementations and reduces the memory overhead caused by
temporary variables.

It should be noted that \libsharp\ supports hybrid MPI and OpenMP
parallelisation, which allows, e.g., running an SHT on eight nodes with four
CPU cores each, by specifying eight MPI tasks, each of them consisting of
four OpenMP threads. In general, OpenMP should be preferred over MPI
whenever shared memory is available (i.e.\ at the computing node level), since
the OpenMP algorithms contain dynamic load balancing and have a smaller
communication overhead.

\subsection{Map synthesis of first derivatives}
Generating maps of first derivatives from a set of \alm\ is closely related
to performing an SHT of spin 1. A specialised SHT mode was added to
\libsharp\ for this purpose; it takes as input a set of spin-0 \alm\ and
produces two maps containing $\partial f/\partial \vartheta$ and
$\partial f/(\partial \varphi \sin\vartheta)$, respectively.

\subsection {Support for additional spherical grids}
\label{gridtypes}

Direct support for certain classes of spherical grids has been extended
in comparison to \libpsht; these additions are listed below in detail.
It must be stressed, however, that \libsharp\ can -- very much as
\libpsht\ does -- perform SHTs on any iso-latitude grid with equidistant
pixels on each ring. This very general class of pixelisations includes, e.g.,
certain types of partial spherical coverage. For these general grids, however,
the user is responsible for providing correct quadrature weights when performing a
spherical harmonic analysis.

\subsubsection{Extended support for ECP grids}
\Libpsht\ provides explicit support for HEALPix grids, Gauss-Legendre grids,
and a subset of equidistant cylindrical projection (ECP) grids. The latter are
limited to an even number of rings at the colatitudes
\begin{equation}
\vartheta_n=\frac{(n+0.5)\pi}{N}\text{, }\quad
  n \in [0;N-1]\text{.}
\end{equation}
The associated quadrature weights are given by Fej\'er's first rule
\citep{fejer-1933,gautschi-1967}.

\Libsharp\ extends ECP grid support to allow even and odd numbers of rings,
as well as the colatitude distributions
\begin{equation}
\vartheta_n=\frac{n\pi}{N}\text{, }\quad
  n \in [1;N-1]
\end{equation}
(corresponding to Fej\'er's second rule), and
\begin{equation}
\vartheta_n=\frac{n\pi}{N}\text{, }\quad
  n \in [0;N]
\end{equation}
(corresponding to Clenshaw-Curtis quadrature).
This last pixelisation is identical to the one adopted in
\citet{huffenberger-wandelt-2010}.

Accurate computation of the quadrature weights for these pixelisations is
nontrivial; \libsharp\ adopts the FFT-based approach described in
\citet{waldvogel-2006} for this purpose.

\subsubsection{Reduced Gauss-Legendre grid}
\label{reduced_grid}
The polar optimisation described in Sect.~\ref{polaropt} implies that it is
possible to reduce the number of pixels per ring below the theoretically
required value of $2\lmax+1$ close to the poles.
Eq.~\eqref{eq_polaropt} can be solved for $m$ (at a given $s$, \lmax\ and
$\vartheta$), and using $2m+1$ equidistant pixels in the corresponding map
ring results in a pixelisation that can represent a band-limited function
up to the desired precision, although it is no longer exact in a mathematical sense.
If this number is further increased to the next number composed entirely
of small prime factors (2, 3, and 5 are used in \libsharp's case), this has
the additional advantage of allowing very efficient FFTs.

\Libsharp\ supports this pixel reduction technique in the form of a thinned-out
Gauss-Legendre grid. At moderate to high resolutions ($\lmax>1000$),
more than 30\% of pixels can be saved, which can be significant in various
applications.

It should be noted that working with reduced Gauss-Legendre grids, while saving
considerable amounts of memory, does not change SHT execution times
significantly; all potential savings are already taken into account, for all
grids, by \libsharp's implementation of polar optimisation.

\subsection {Adjoint and real SHTs}
Since Eq. \eqref{syn_used}
is a linear transform, we can introduce the notation
\begin{equation}
  \label{linalg-syn}
  \mathbf{p} = \mathbf{Y} \mathbf{a}
\end{equation}
for a vector $\mathbf{a}$ of spherical harmonic coefficients and
corresponding vector $\mathbf{p}$ of pixels. Similarly, one can write
Eq. \eqref{ana_used} as
\begin{equation}
  \label{linalg-ana}
  \mathbf{a} = \mathbf{Y}^T \mathbf{W} \mathbf{p},
\end{equation}
where $\mathbf{W}$ is a diagonal matrix of quadrature weights.  When
including SHTs as operators in linear systems, one will often need the
\textit{adjoint spherical harmonic synthesis}, $\mathbf{Y}^T$, and the
\textit{adjoint spherical harmonic analysis}, $\mathbf{W}\mathbf{Y}$. For
instance, if $\mathbf{a}$ is a random vector with covariance matrix
$\mathbf{C}$ in the spherical harmonic domain, then its pixel
representation $\mathbf{Y}\mathbf{a}$ has the covariance matrix
$\mathbf{Y}\mathbf{C}\mathbf{Y}^T$. Multiplication by this matrix
requires the use of the adjoint synthesis, which corresponds to analysis
with a different choice of weights. \Libsharp{} includes routines for
adjoint SHTs, which is more user-friendly than having to compensate
for the wrong choice of weights in user code, and also avoids an extra
pass over the data.

For linear algebra computations, the vector $\mathbf{a}$ must also
include $a_{lm}$ with $m<0$, even if \libsharp{} will only compute the
coefficients for $m \ge 0$.  The use of the \textit{real spherical
harmonics} convention is a convenient way to include negative $m$
without increasing the computational workload by duplicating all
coefficients. For the definition we refer to the appendix of
\cite{oliviera-costa2004}.  The convention is supported directly in
\libsharp{}, although with a restriction in the storage scheme: The
coefficients for $m<0$ must be stored in the same locations as the
corresponding imaginary parts of the complex coefficients, so that the
pattern in memory is $[a_{l,m}, a_{l,-m}]$.


\section{Evaluation}

Most tests were performed on the SuperMUC Petascale System located at
the Leibniz-Rechenzentrum Garching. This system consists of nodes
containing 32GB of main memory and 16 Intel Xeon E5-2680 cores running
at 2.7GHz.  The exception is the comparison with Wavemoth, which was
performed on the Abel cluster at the University of Oslo on very
similar hardware; Xeon E5-2670 at 2.6 GHz.

The code was compiled with \texttt{gcc} version 4.7.2. The Intel compiler
(version 12.1.6) was also tested, but produced slightly inferior code.

Except where noted otherwise, test calculations were performed using the
reduced Gauss-Legendre grid (see Sect.~\ref{reduced_grid}) to represent
spherical map data. This was done for the pragmatic purpose of minimising
the tests' memory usage, which allowed going to higher band limits in some
cases, as well as to demonstrate the viability of this pixelisation.

The band limits adopted for the tests all obey $\lmax=2^n-1$ with
$n\in \mathbb{N}$ (except for those presented in Sect.~\ref{weakscaling}).
This is done in analogy to most other papers on the subject,
but leads to some unfamiliar numbers especially at very high \lmax.

The number of cores used for any particular run always is a power of 2.

\subsection{Accuracy tests}

\subsubsection {Comparison with other implementations}
The numerical equivalence of \libsharp's SHTs to existing implementations was
verified by running spherical harmonic synthesis transforms on a Gauss-Legendre
grid at $\lmax=50$ and spins 0, 1, and 2 with both \libsharp\ and \libpsht,
and comparing the results. The differences of the results lay well within the
expected levels of numerical errors caused by the finite precision of IEEE
numbers. Spherical harmonic analysis is implicitly tested by the experiments
in the following sections.

\subsubsection {Evaluation of SHT pairs}
\label{acctest}
\begin{figure}
\centering
\includegraphics[width=0.97\columnwidth]{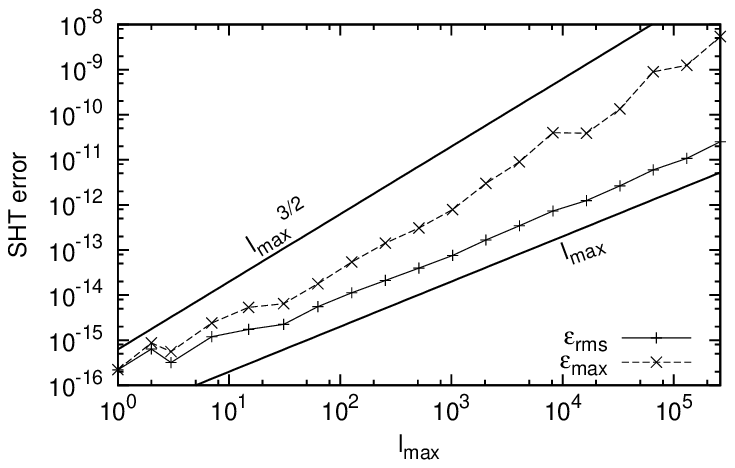}
\caption{Maximum and rms errors for inverse/forward spin=0 SHT pairs at different
\lmax.}
\label{fig_accuracy}
\end{figure}

To test the accuracy of \libsharp's transforms, sets of spin=0 \alm\ coefficients
were generated by setting their real and imaginary parts to numbers drawn from
a uniform random distribution in the range [-1;1[ (with exception of the
imaginary parts for $m=0$, which have to be zero for symmetry reasons).
This data set was transformed onto a reduced Gauss-Legendre grid and back to
spherical harmonic space again, resulting in $\hat a_{lm}$.

The rms and maximum errors of this inverse/forward transform pair can be
written as
\begin{align}
\varepsilon_{\text{rms}} &:= \sqrt{\frac{\sum_{lm}|{}_sa_{lm}-{}_s\hat a_{lm}|^2}{\sum_{lm}|{}_sa_{lm}|^2}}\quad\text{and} \\
\varepsilon_{\text{max}} &:= \max_{lm} \left| {}_sa_{lm}-{}_s\hat a_{lm}\right|\text{.}
\end{align}

Fig.~\ref{fig_accuracy} shows the measured errors for a wide range of band
limits. As expected, the numbers are close to the accuracy limit of double
precision IEEE numbers for low \lmax; rms errors increase roughly linearly with
the band limit, while the maximum error seems to exhibit an $\lmax^{3/2}$ scaling.
Even at $\lmax=262143$ (which is extremely high compared to
values typically required in cosmology), the errors are still negligible
compared with the uncertainties in the input data in today's experiments.

Analogous experiments were performed for spins 2 and 37, with very similar
results (not shown).

\subsection{Performance tests}

Determining reliable execution times for SHTs is nontrivial at low band limits,
since intermittent operating system activity can significantly distort the
measurement of short time scales.
All \libsharp\ timings shown in the following sections were obtained using the
following procedure: the SHT pair in question is executed repeatedly until the
accumulated wall-clock time exceeds 2 seconds. Then the shortest measured
wall-clock time for synthesis and analysis is selected from the available set.

\subsubsection{Strong scaling test}
\begin{figure}
\centering
\includegraphics[width=0.97\columnwidth]{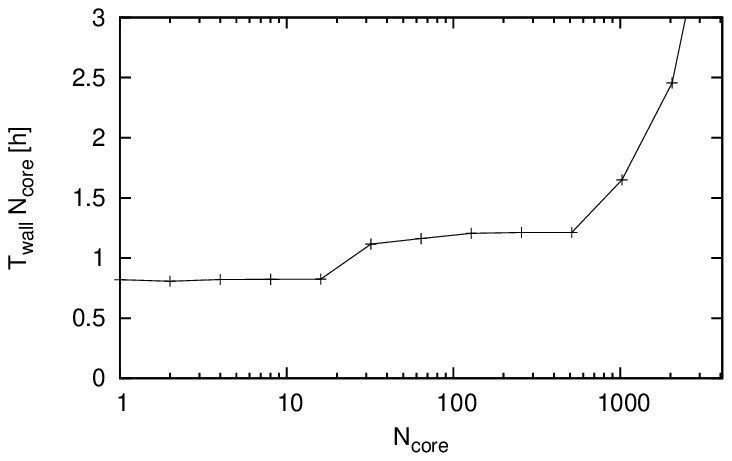}
\caption{Strong scaling scenario: accumulated wall-clock time for a spin=2
SHT pair with \lmax=16383 run on various numbers of cores.}
\label{fig_strongscaling}
\end{figure}
To assess strong scaling behaviour (i.e.\ run time scaling for a given problem
with fixed total workload), a spin=2 SHT with \lmax=16383 was carried
out with differing degrees of parallelisation. The accumulated wall-clock time
of these transforms (synthesis + analysis) is shown in
Fig.~\ref{fig_strongscaling}. It is evident
that the scaling is nearly ideal up to 16 cores, which implies that
parallelisation overhead is negligible in this range. Beyond 16 cores, MPI
communication has to be used for inter-node communication, and this most likely
accounts for the sudden jump in accumulated time. At even higher core counts,
linear scaling is again reached, although with a poorer proportionality factor
than in the intra-node case. Finally, for 1024 and more cores, the communication
time dominates the actual computation, and scalability is lost.

\subsubsection{Weak scaling test}
\label{weakscaling}
\begin{figure}
\centering
\includegraphics[width=0.97\columnwidth]{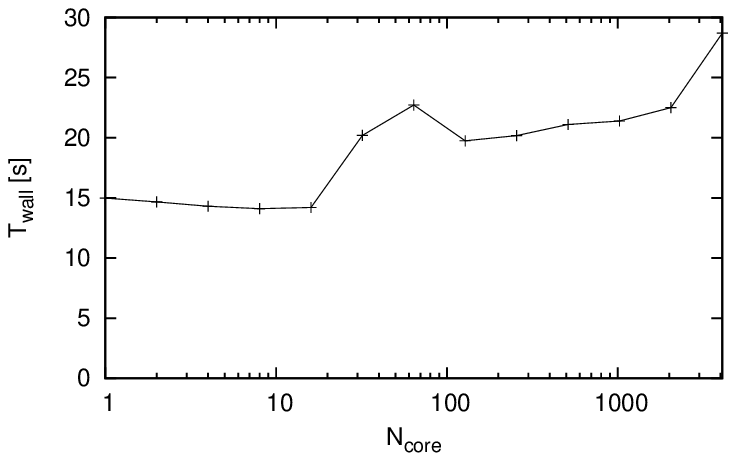}
\caption{Weak scaling scenario: wall-clock time for a spin=0 SHT pair run on
various numbers of cores, with constant amount of work per core
($\lmax(N)=4096\cdot N^{1/3}-1$).}
\label{fig_weakscaling}
\end{figure}
Weak scaling behaviour of the algorithm is investigated by choosing problem
sizes that keep the total work \textit{per core} constant, in contrast
to a fixed \textit{total} workload. Assuming an SHT complexity of
\compl{\lmax^3}, the band limits were derived from the employed number of cores $N$ via
$\lmax(N)=4096\cdot N^{1/3}-1$. The results are shown in Fig.~\ref{fig_weakscaling}.
Ideal scaling corresponds to a horizontal line. Again, the transition from
one to several computing nodes degrades performance, whereas scaling on a single
node, as well as in the multi-node range, is very good.
By keeping the amount of work per core constant, the breakdown of scalability
is shifted to 4096 cores, compared with 1024 in the strong scaling test.

It is interesting to note that the scaling within a single node is actually
slightly superlinear; this is most likely because in this setup,
the amount of memory per core decreases with increasing problem size, which
in turn can improve the amount of cache re-use and reduce memory bandwidth per core.

\subsubsection{General scaling and efficiency}
\begin{figure}
\centering
\includegraphics[width=0.97\columnwidth]{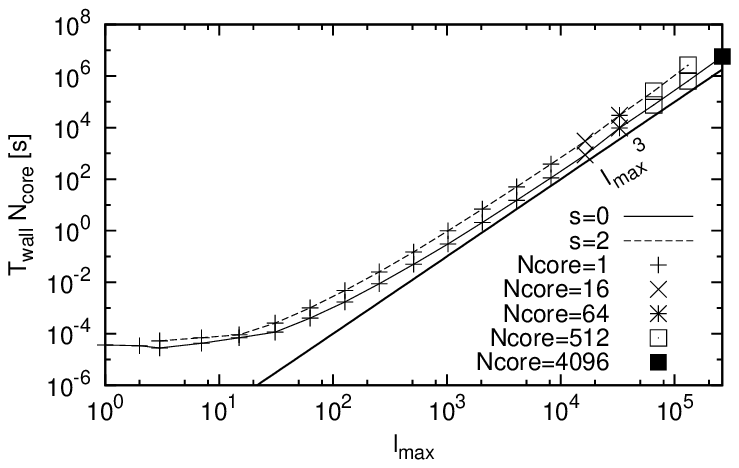}
\caption{Accumulated wall-clock time for spin=0 and spin=2 SHT pairs at a wide range
of different band limits. For every run the number of cores was chosen
sufficiently small to keep parallelisation overhead low.}
\label{fig_runtime}
\end{figure}
The preceding two sections did not cover cases with small SHTs. This scenario
is interesting, however, since in the limit of small \lmax\ those components
of the SHT implementation with complexities lower than \compl{\lmax^3}
(like the FFT steps of Eqs.~\ref{syn1} and \ref{ana2}) may begin
to dominate execution time. Fig.~\ref{fig_runtime} shows the total wall-clock
time for SHT pairs over a very wide range of band limits; to minimise the impact
of communication, the degree of parallelisation was kept as low as possible
for the runs in question. As expected, the $\lmax^3$ scaling is a very good
model for the execution times at $\lmax\geq511$. Below this limit, the FFTs,
precomputations for the spherical harmonic recursion, memory copy operations
and other parts of the code begin to dominate.

\begin{figure}
\centering
\includegraphics[width=0.97\columnwidth]{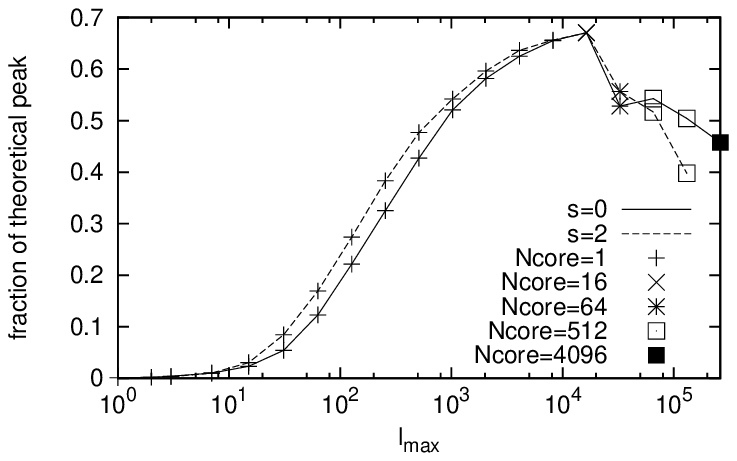}
\caption{Fraction of theoretical peak performance reached by various SHT pairs.
For every run the number of cores was chosen sufficiently small to keep
parallelisation overhead low.
}
\label{fig_percentpeak}
\end{figure}
In analogy to one of the tests described in \citet{reinecke-2011}, we computed
a lower limit for the number of executed floating-point operations per
second in \libsharp's SHTs and compared the result with the theoretical peak
performance achievable on the given hardware, which is eight operations per clock
cycle (four additions and four multiplications) or 21.6\,GFlops/s per core.
Fig.~\ref{fig_percentpeak} shows the results. In contrast to \libpsht, which
reached approximately 22\% for $s=0$ and 43\% for $s=2$, both scalar and
tensor harmonic transforms exhibit very similar performance levels and almost
reach 70\% of theoretical peak in the most favourable regime, thanks to the
changed structure of the inner loops. For the
\lmax\ range that is typically required in cosmological applications,
performance exceeds 50\% (even when MPI is used), which is very high for a
practically useful algorithm on this kind of computer architecture.

\subsubsection{Multiple simultaneous SHTs}
\begin{figure}
\centering
\includegraphics[width=0.97\columnwidth]{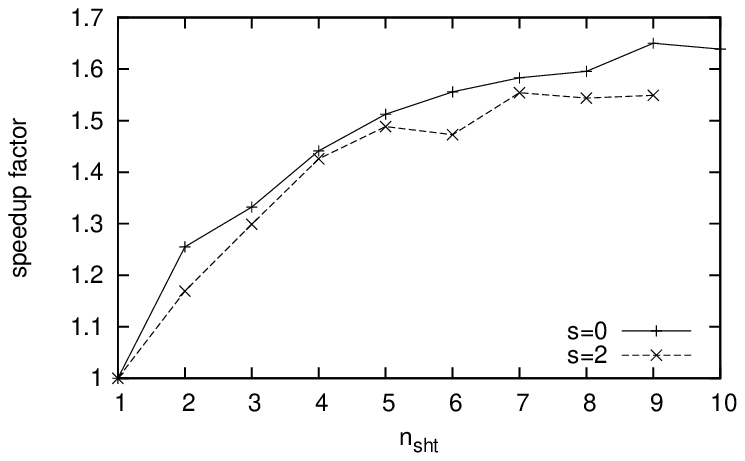}
\caption{Relative speed-up when performing several SHTs simultaneously, compared
with sequential execution. The SHT had a band limit of $\lmax=8191$. }
\label{fig_multi_sht}
\end{figure}
The computation of the \lamlm\ coefficients accounts for roughly half the
arithmetic operations in an SHT. If several SHTs with identical grid geometry
and band limit are computed simultaneously, it is possible to re-use these
coefficients, thereby reducing the overall operation count.
Fig.~\ref{fig_multi_sht} shows the speed-ups compared to sequential execution
for various scenarios, which increase with the number of transforms and reach
saturation around a factor of 1.6. This value is lower than the na\"ively
expected asymptotic factor of 2 (corresponding to avoiding half of the
arithmetic operations), since the changed algorithm requires more memory
transfers between Level-1 cache and CPU registers and therefore operates at
a lower percentage of theoretical peak performance. Nevertheless, running
SHTs simultaneously is evidently beneficial and should be used whenever
possible.

\subsubsection{Memory overhead}
\begin{figure}
\centering
\includegraphics[width=0.97\columnwidth]{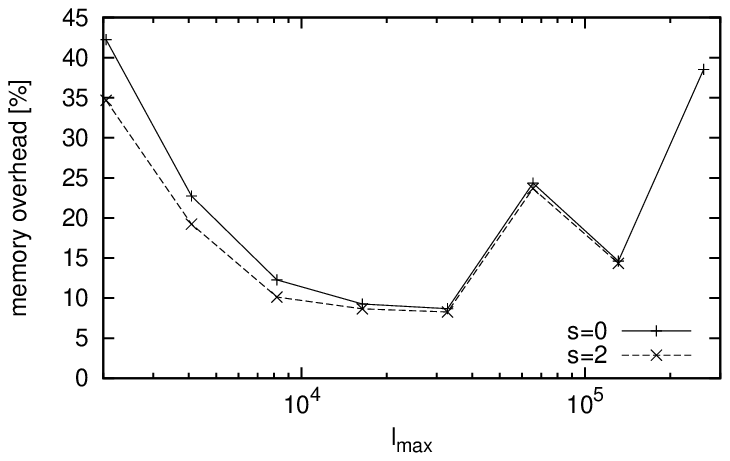}
\caption{Relative memory overhead, i.e.\ the fraction of total memory that is
not occupied by input
and output data of the SHT. For low \lmax\ this is dominated by the program
binary, for high \lmax\ by temporary arrays.}
\label{fig_mem_oh}
\end{figure}
Especially at high band limits, it is important that the SHT library does not
consume a large amount of main memory, to avoid memory exhaustion and subsequent
swapping or code crashes. \Libsharp\ is designed with the goal to keep the size
of its auxiliary data structures much lower than the combined size of any
SHT's input and output arrays. A measurement is shown in Fig.~\ref{fig_mem_oh}.
Below the lowest shown band limit of 2047, memory overhead quickly climbs to
almost 100\%, since in this regime memory consumption is dominated by the
executable and the constant overhead of the communication libraries, which
on the testing machine amounts to approximately 50MB.
In the important range ($\lmax\geq 2047$), memory overhead lies below 45\%.

\subsubsection{Comparison with existing implementations}
\begin{table*}
\caption{Performance comparison with other implementations at $\lmax=2047$,
$n_\text{core}=1$.}
\begin{center}
\begin{tabular}{lllccccc}
\hline\hline
Code & version & grid & spin & $n_\text{SHT}$ & $R_\text{AVX}$ & $R_\text{SSE2}$ & $R_\text{scalar}$\\
\hline\hline
\shtns                 & 2.31     & Gauss-Legendre        & 0 & 1 & \pnull0.84 & \pnull0.88 & \pnull0.91 \\
Wavemoth (brute-force) & Nov 2011 & HEALPix (\nside=1024) & 0 & 1 & \pnull1.63 & \pnull0.98 & --- \\
''                     & ''       & ''                    & 0 & 5 & \pnull1.59 & \pnull1.09 & --- \\
Wavemoth (butterfly)   & Nov 2011 & HEALPix (\nside=1024) & 0 & 5 & \pnull0.96 & \pnull0.66 & --- \\
\libpsht           & Jan 2011 & Gauss-Legendre        & 0 & 1 & \pnull4.06 & \pnull2.30 & \pnull2.30 \\
''                 & ''       & ''                    & 0 & 5 & \pnull2.66 & \pnull1.75 & \pnull1.62 \\
''                 & ''       & ''                    & 2 & 1 & \pnull2.50 & \pnull1.48 & \pnull1.20 \\
''                 & ''       & ''                    & 2 & 5 & \pnull2.15 & \pnull1.44 & \pnull1.08 \\
\texttt{spinsfast} & r104     & ECP (Clenshaw-Curtis) & 0 & 1 & 57.04 & 32.12 & 15.31 \\
''                 & ''       & ''                    & 0 & 5 & 28.39 & 18.72 & \pnull9.38 \\
''                 & ''       & ''                    & 2 & 1 & 16.99 & 10.20 & \pnull4.73 \\
''                 & ''       & ''                    & 2 & 5 & \pnull8.60 & \pnull5.66 & \pnull2.56 \\
SSHT               & 1.0b1    & MW sampling theorem   & 0 & 1 & 20.91 & 15.60 & \pnull9.46 \\
''                 & ''       & ''                    & 2 & 1 & 13.40 & \pnull9.29 & \pnull4.99 \\
\stwohat           & 2.55beta & HEALPix (\nside=1024) & 0 & 1 & 12.33 & \pnull7.33 & \pnull3.60 \\
Glesp              & 2        & Gauss-Legendre        & 0 & 1 & 55.32 & 31.26 & 14.95 \\
\end{tabular}
\end{center}
\tablefoot{All tests had a band limit of $\lmax=2047$ and were carried
  out on a single core. The grids used by \libsharp\ and the respective comparison
  code were identical in each run. $R_\text{AVX}$ denotes the quotient of wall-clock
  times for the respective code and \libsharp\ in the presence of the AVX
  instruction set, $R_\text{SSE2}$ is the quotient when SSE2 (but not AVX) is
  supported, and $R_\text{scalar}$ was measured with both SSE2 and AVX disabled.
  The \libsharp\ support for the MW sampling
  theorem used for the SSHT comparisons is experimental. For Wavemoth,
  butterfly matrix compression can optionally be enabled. In the benchmark given we
  requested an accuracy of $10^{-4}$, which led to an extra requirement of 4 GB of
  precomputed data in memory. Note that when running on a single core, Wavemoth is
  at an advantage compared to the normal situation where the memory bus is shared
  between multiple cores.}
\label{perfnumbers}
\end{table*}
Table~\ref{perfnumbers} shows a performance comparison of synthesis/analysis
SHT pairs between \libsharp\ and various other SHT implementations. In addition
to the already mentioned \shtns, Wavemoth, \stwohat\ and \libpsht\ codes, we
also included {\tt spinsfast} \citep{huffenberger-wandelt-2010},
SSHT \citep{mcewen-wiaux-2011} and Glesp \citep{doroshkevich-etal-2005} in the
comparison. All computations
shared a common band limit of 2047 and were executed on a single
core, since the corresponding SHTs are supported by all libraries and are very likely
carried out with a comparatively high efficiency by all of them. The large
overall number of possible parameters (\lmax, spin, number of simultaneous
transforms, degree and kind of parallelisation, choice of grid, etc.) prevented
a truly comprehensive study.

Overall, \libsharp's performance is very satisfactory and exhibits speed-ups
of more than an order of magnitude in several cases. The table also
demonstrates \libsharp's flexibility, since it supports all of the other codes'
``native'' grid geometries, which is required for direct comparisons.

The three last columns list time ratios measured under different assumptions:
$R_\text{AVX}$ reflects values that can be expected on modern (2012 and later)
AMD/Intel CPUs supporting AVX, $R_\text{SSE2}$ applies to older (2001 and later)
CPUs with the SSE2 instruction set. $R_\text{scalar}$ should be used for CPUs
from other vendors like IBM or ARM, since \libsharp\ does not yet support
vectorisation for these architectures.

\begin{figure}
\centering
\includegraphics[width=0.97\columnwidth]{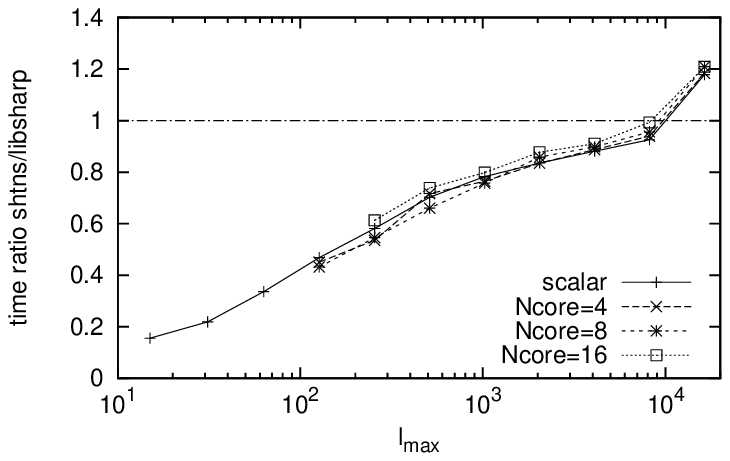}
\caption{Performance comparison between \libsharp\ and \shtns\ for
varying \lmax\ and number of OpenMP threads. Note that reduced
autotuning was used for \shtns\ at \lmax=16383 (see text).}
\label{fig_shtns_comp}
\end{figure}
Fig.~\ref{fig_shtns_comp} shows the relative performance of identical
SHT pairs on a full Gauss-Legendre grid with s=0 for \libsharp\ and
\shtns. For these measurements the benchmarking code delivered with \shtns\ was
adjusted to measure SHT times in a similar fashion as was described above.
The plotted quantity is \shtns\ wall-clock time divided
by \libsharp\ wall-clock time for varying \lmax\ and number of OpenMP threads.
It is evident that \shtns\ has a
significant advantage for small band limits (almost an order of magnitude)
and maintains a slight edge up to \lmax=8191.
It must be noted, however, that the measured times do not
include the overhead for auto-tuning and necessary precalculations, which
in the case of \shtns\ are about an order of magnitude more expensive
than the SHTs themselves. As a consequence, its performance advantage only
pays off if many identical SHT operations are performed within one run.
The origin of \shtns's performance advantage has not been studied in depth;
however, a quick analysis shows that the measured time differences scale
roughly like $\lmax^2$, so the following explanations are likely candidates:
\begin{itemize}
\item \libsharp\ performs all of its precomputations as part of the time-measured
  SHT
\item \libsharp's flexibility with regard to pixelisation and storage arrangement
  of input and output data requires some additional copy operations
\item at low band limits the inferior performance of \libsharp's FFT
  implementation has a noticeable impact on overall run times.
\end{itemize}

The relative performance of both libraries is remarkably insensitive
to the number of OpenMP threads; this indicates that the performance
differences are located in parallel code regions as opposed to sequential ones.

For $\lmax=16383$, the time required by the default \shtns\ autotuner
becomes very long (on the order of wall-clock hours), so that we decided
to invoke it with an option for reduced tuning. It is likely
a consequence of this missed optimisation that, at this band limit,
\libsharp\ is the better-performing code.

\section{Conclusions}
Judging from the benchmarks presented in the preceding section, the goals
that were set for the \libsharp\ library have been reached: it exceeds
\libpsht\ in terms of performance, supports recent developments in
microprocessor technology, allows using distributed memory systems for a wider
range of applications, and is slightly easier to use. On the developer side,
the modular design of the code makes it much more straightforward to add support
for new instruction sets and other functionality.

In some specific scenarios, especially for SHTs with comparatively low band
limits, \libsharp\ does not provide the best performance of all available
implementations, but given its extreme flexibility concerning grid types and
the memory layout of its input/output data, as well as its compactness
($\approx8000$ lines of portable and easily maintainable source code without
external dependencies), this compromise certainly seems acceptable.

The library has been successfully integrated into version 3.1 of the HEALPix
C++ and Fortran packages. There also exists an experimental version of the
SSHT\fnurl{http://www.mrao.cam.ac.uk/~jdm57/ssht/index.html} package with
\libsharp\ replacing the library's original SHT engine.
\Libsharp\ is also used as SHT engine in an upcoming version of the
Python package \nifty\fnurl{http://www.mpa-garching.mpg.de/ift/nifty/}
for signal inference \citep{selig-etal-2013}. Recently, the total
convolution code \texttt{conviqt}\ \citep{prezeau-reinecke-2010}, which is
a central component of the \textit{Planck} simulation pipeline
\citep{reinecke-etal-2006}, has been updated and is now based on \libsharp\ SHTs.
There are plans for a similar update of the \texttt{artDeco} deconvolution map
maker \citep{keihanen-reinecke-2012}.

A potential future field of work is porting \libsharp\ to Intel's
``many integrated cores''
architecture\fnurl{http://en.wikipedia.org/wiki/Intel_MIC}, once sufficient
compiler support for this platform has been established. The hardware appears
to be very well suited for running SHTs, and the porting by itself would provide
a welcome test for the adaptability of the library's code design.

\begin{acknowledgements}
We thank our referee Nathana\"el Schaeffer for his constructive remarks
and especially for pointing out a missed optimisation opportunity in our
\shtns\ installation, which had a significant effect on some benchmark results.
MR is supported by the German Aeronautics Center and Space Agency (DLR), under
program 50-OP-0901, funded by the Federal Ministry of Economics and
Technology. DSS is supported by the European Research Council, grant StG2010-257080.
The presented benchmarks were performed as project \textit{pr89yi}
at the Leibniz Computing Center Garching.
\end{acknowledgements}

\bibliographystyle{aa_arxiv}
\bibliography{planck}

\end {document}